\documentclass[]{pasj01}
\usepackage[switch,mathlines]{lineno}

\Received{2024/10/14}
\Accepted{2025/5/21}
 
\begin{document} 

\title{ 
Magnetic fields in the massive star-forming region NGC~6334 and their relationship with the properties of dust filaments probed by [C\,\emissiontype{II}] and PAH emissions

} 

\author{Takayoshi \textsc{Kusune}\altaffilmark{1}%
}
\altaffiltext{1}{Graduate School of Science, Nagoya University, Furo-cho, Chikusa-ku, Nagoya, Aichi 464-8602, Japan}
\email{takayoshi.kusune@u.phys.nagoya-u.ac.jp}

\author{Hayata \textsc{Tsuji},\altaffilmark{1}}

\author{Shinki \textsc{Oyabu}\altaffilmark{2}}
\altaffiltext{2}{Institute of Liberal Arts and Sciences, Tokushima University, Minami Jousanajima-machi 1-1, Tokushima 770-8502, Japan} 

\author{Hidehiro \textsc{Kaneda}\altaffilmark{1}} 

\author{Toyoaki \textsc{Suzuki}\altaffilmark{3}} 
\altaffiltext{3}{Institute of Space and Astronautical Science, Japan Aerospace Exploration Agency 3-1-1 Yoshinodai, Chuo-ku, Sagamihara, Kanagawa 252-5210, Japan} 

\author{Akiko \textsc{Yasuda}\altaffilmark{1}} 

\author{Devendra \textsc{Ojha}\altaffilmark{4}} 
\altaffiltext{4}{Tata Institute of Fundamental Research, Homi Bhabha Road, Colaba, Mumbai 400005, India} 

\author{Swarna K. \textsc{Ghosh}\altaffilmark{4}}

\author{Koshvendra \textsc{Singh}\altaffilmark{4}}

\author{Joe P. \textsc{Ninan}\altaffilmark{4}}



\maketitle

\begin{abstract}
We carried out the near-infrared ($JHK_{\rm s}$) imaging polarimetric observation with the polarimeter SIRPOL on the Infrared Survey Facility (IRSF) 1.4 m telescope
and [C\,\emissiontype{II}] line mapping observation with a Fabry-P\'{e}rot spectrometer on board a 100-cm TIFR balloon-borne far-infrared telescope toward NGC~6334, 
and revealed the relationship between the plane-of-sky (POS) magnetic fields and [C\,\emissiontype{II}] emission lines 
to investigate the star formation in the molecular cloud. 
The polarization vector map shows that 
the POS magnetic fields are approximately perpendicular to the main filament elongation of NGC~6334. 
On the other hand, 
the POS magnetic fields tend to be parallel or random for the other filaments in NGC~6334. 
The [C\,\emissiontype{II}] emission shows a distribution well aligned with the main filament. 
Strong [C\,\emissiontype{II}] emission is also seen in the hub-filament system. 
Since the main filament is sandwiched between two H\,\emissiontype{II} regions, 
it is most likely that gas is efficiently accreting from the shells of the H\,\emissiontype{II} regions along the magnetic field 
resulting in active star formation. 
This is consistent with the NGC~6334 being bright in [C\,\emissiontype{II}] emission. 
\end{abstract}

\section{Introduction} 


In the 2010s, 
observations of thermal emission from the dust and of molecular gas have revealed the universal presence of dense filamentary structures in the molecular clouds 
(e.g., \cite{andre10,molinari10}). 
Filaments rarely exist in isolation, 
and a structure called as a hub-filament system is well-known,
where multiple filaments appear to merge. 
In the hub region, 
the density is higher than that of the surrounding filaments, 
and stellar clusters and high-mass stars are often associated. 
Therefore, 
the hub-filament system in the molecular cloud may be one of the key steps in the high-mass star formation process (e.g., \cite{myers09,schneider12,peretto13,kumar22}).  
Magnetic fields are believed to play an important role in the formation and maintenance of filamentary structures,  
as well as in the star formation process within them; 
however, their behavior has not yet been fully understood. 

Wide-field near-infrared (NIR) polarimetry is a good tool 
for measuring the plane-of-sky (POS) magnetic field direction toward molecular clouds (e.g., \cite{tamura07}); 
we can determine the direction of the POS magnetic field, 
by measurement of the linear polarization produced by dust grains aligned with magnetic field 
in the dichroic extinction of background starlight.

The [C\,\emissiontype{II}] 158 $\mu$m emission line is the most dominant cooling line in the photo-dissociation regions (PDRs) associated with high-mass stars,
and thus is an excellent tracer of a high-mass star-forming region in a dense molecular cloud (e.g., \cite{ht99}). 
Additionally, 
the [C\,\emissiontype{II}] emission line is known as the brightest far-infrared line in star-forming galaxies 
and has been widely used in observations of extragalaxy (e.g., \cite{stacey10,madden20}). 
Note that 
in our balloon-borne observations, 
due to the low sensitivity of the detector, 
the [C\,\emissiontype{II}] emission does not trace non-star forming gas. 
 
\begin{figure*}
 \begin{center}
  \includegraphics[width=15cm]{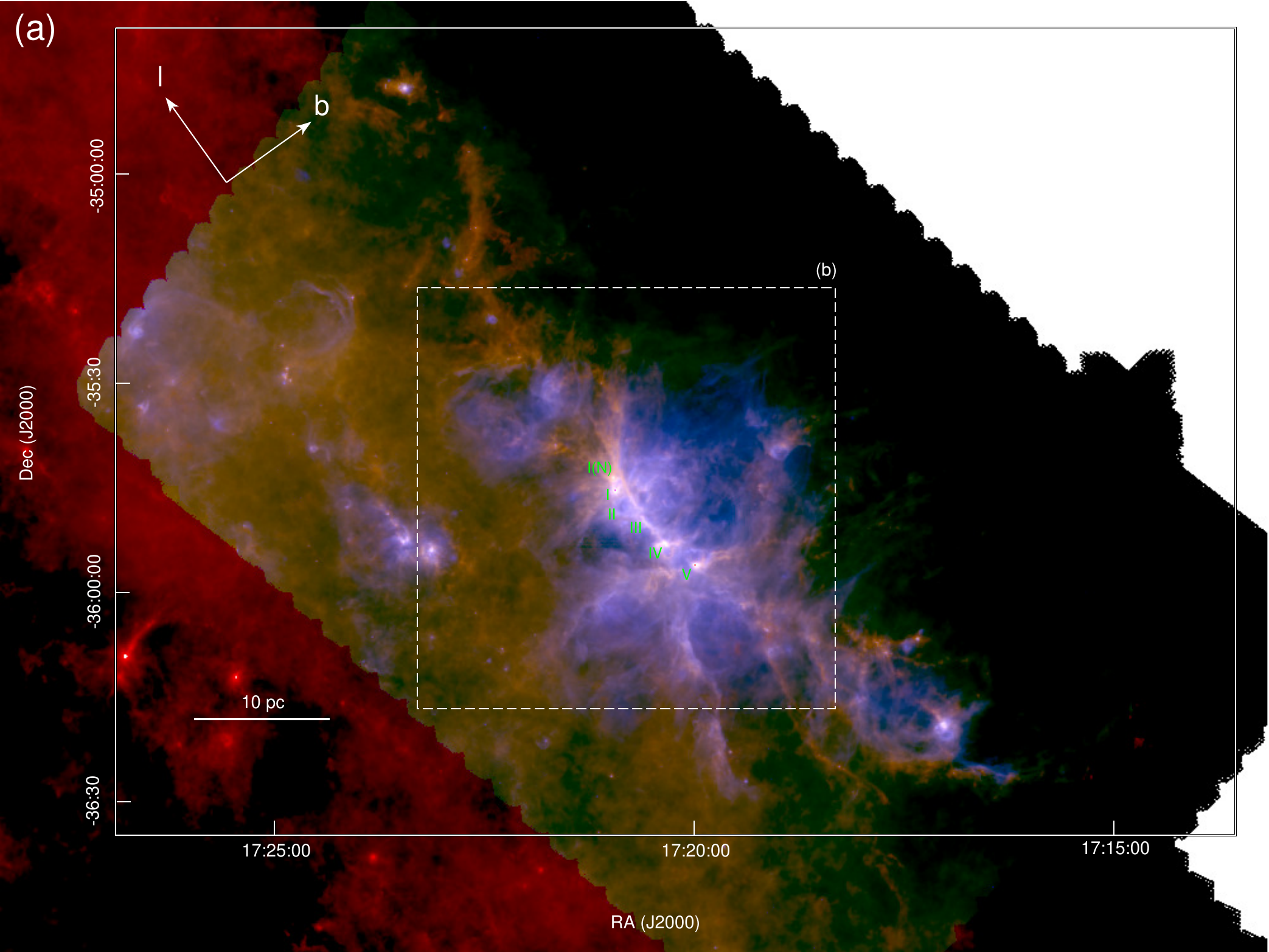}
  \includegraphics[width=9cm]{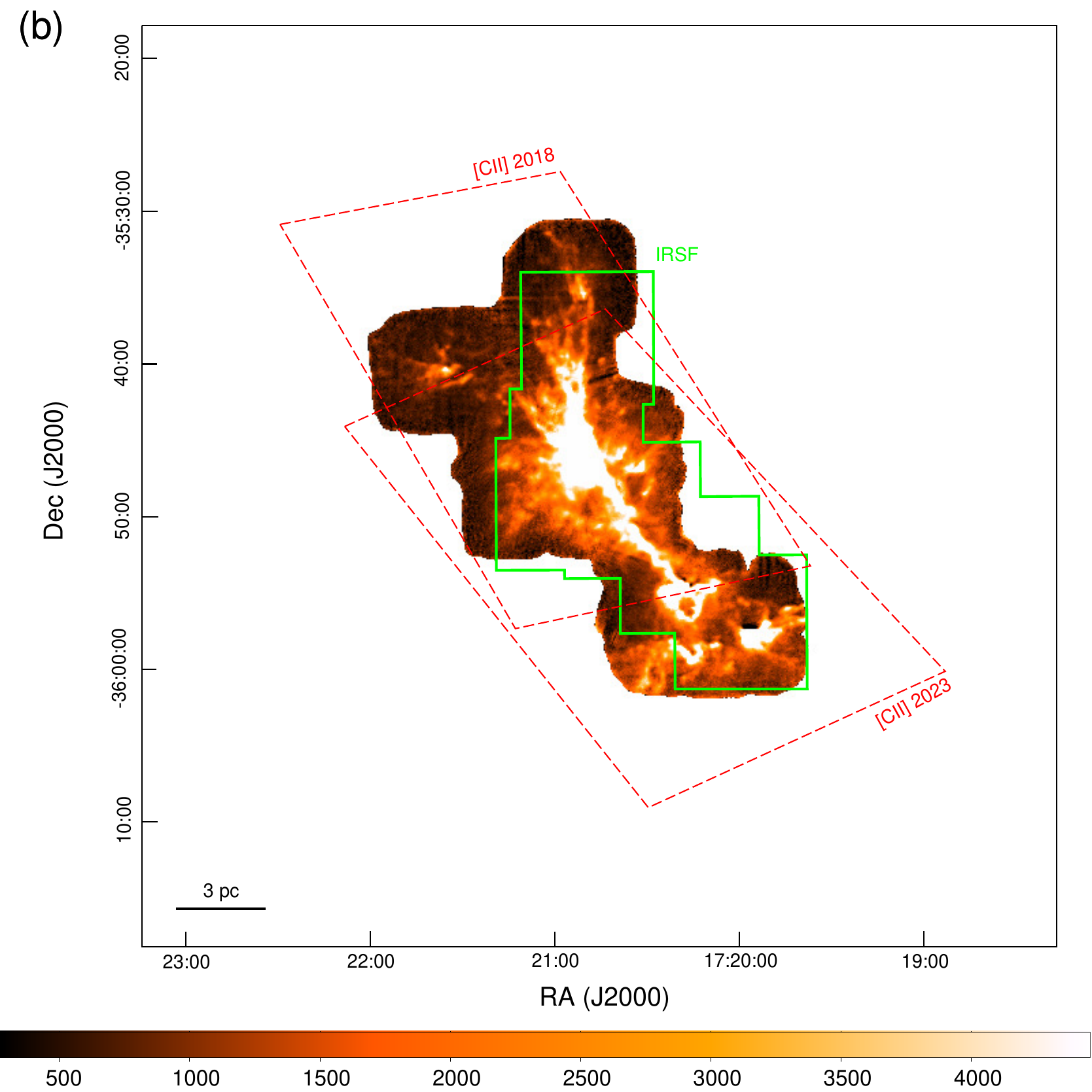}  
 \end{center} 
\caption{ 
(a) Three-color composite Herschel image of the NGC~6334 region (250 $\mu$m in red; 160 $\mu$m in green; 70 $\mu$m in blue). 
Six active star-forming regions are labeled in green. 
The white dashed square indicates the area in figure 1b.
(b) ArT\'{e}MiS 350 $\mu$m dust continuum map from \citet{andre16} toward NGC~6334 in unit of MJy sr$^{-1}$. 
The areas of our NIR polarimetric observations with the IRSF 1.4 m telescope and of our [C\,\emissiontype{II}] observations 
are indicated by solid green polygon and dashed red boxes, respectively. 
{Alt text: Two maps showing the overall view of NGC 6334 and indicating the area we observed.}
}
\label{obs_area}
\end{figure*}

NGC~6334 is one of the most prominent Galactic massive star-forming regions located at a distance of $1.76$ kpc \citep{russeil20}. 
Figure \ref{obs_area}a shows an overview of NGC~6334 using the three-color composite Herschel image. 
This object has an elongated structure parallel to the Galactic plane, 
with a length of about 50 pc, 
in which in the central dense gas region of NGC~6334, 
there is a prominent filamentary structure extending from north to south, with six massive star-forming molecular clumps associated along the filaments. 
In these clumps, active intermediate- or high-mass star formations are known to be undergoing 
(e.g., \cite{andre16, russeil12}). 
NGC~6334 has been observed and studied at various wavelengths (e.g., \cite{persi08}). 
A number of H\,\emissiontype{II} regions are associated with NGC~6334, 
which is also known as the ``Cats Paw Nebula'' because of its appearance. 
As shown in figure \ref{obs_area}, 
the prominent filamentary structure of $\sim$10 pc in length is located between these H\,\emissiontype{II} regions (e.g., \cite{shimajiri19}). 
This filamentary cloud has a dense ridge in the center and a hub structure at the northeast end. 
\citet{fukui18} carried out CO observations with NANTEN2 and ASTE toward NGC~6334, 
and mentioned that 
the cloud-cloud collision over a 100 pc lead the active star formation in NGC~6334.

There are some observational papers on the magnetic field of NGC~6334 at various spatial scales. 
\citet{li15} revealed that 
the directions of magnetic fields do not change much over the entire scale range, 
meaning that the magnetic fields are dynamically important for the gravitational collapse of the molecular cloud. 
\citet{arzou21} conducted the far-infrared polarimetric observation with the James Clerk Maxwell Telescope (JCMT), 
and showed its magnetic field structures of filaments in detail. 
They found that 
in the sub-filaments of the cloud, the magnetic field is perpendicular or random to the crest, 
and that where the sub-filament connects to the ridge or the hubs the magnetic field is parallel to the crests, 
which indicated the flows of infalling gas from sub-filaments to main filament. 
\citet{li23} conducted the ALMA dust polarimetric observations toward four massive clumps in NGC~6334 
and combined them with the large-scale observations such as JCMT, Planck, and NANTEN2  
to reveal the relative orientation of the magnetic fields, gas column density gradients, local gravity, and velocity gradients.

For NGC~6334,
there have been some papers reporting its magnetic field properties, 
but none have observed [C\,\emissiontype{II}] emission line.  
So far, 
observations of magnetic fields and the [C\,\emissiontype{II}] emission in a specific region have been limited to the high-mass star-forming region RCW~36 \citep{bji24}.
No such observations have been reported for NGC 6334, as presented in the present paper. 
 
In the present paper, 
we report the results of NIR polarimetric observation and the [C\,\emissiontype{II}] line observation toward NGC~6334,
and 
investigate the relationship between the magnetic field and filament structure, and between the [C\,\emissiontype{II}] emission and the PAH emission around NGC~6334. 
In section 2, we describe the details of NIR polarimetric observation, the far-infrared [C\,\emissiontype{II}] line observation, and Herschel archive data.
In section 3, we describe the dust distribution, magnetic field structure, and the [C\,\emissiontype{II}] distribution of NGC~6334. 
In section 4, we discuss the massive star formation in NGC~6334 mainly in terms of the magnetic fields, 
and compare the [C\,\emissiontype{II}] emission with the PAH emission. 
We give a summary of the analysis and results in section 5.

\section{Observations and data analysis}

\subsection{Near-infrared polarimetric observations with IRSF} 

NIR polarimetric observations of NGC~6334 were conducted 
with the Infrared Survey Facility (IRSF) 1.4~m telescope at the South African Astronomical Observatory. 
We used the imaging polarimeter SIRPOL \citep{kandori06} 
that is a polarimetry mode of the SIRIUS camera 
equipped with three $1024\times1024$ HdCdTe (HAWAII) arrays, $JHK_{\rm s}$ filters, and dichroic mirrors, 
which enable simultaneous $JHK_{\rm s}$ observations (\cite{nagashima99}; \cite{nagayama03}). 
The field of view at each band is $\sim$7$\farcm7\times7\farcm7$ with a pixel scale of 0\farcs45. 
We used the pyIRSF pipeline software\footnote{https://sourceforge.net/projects/irsfsoftware/} 
to create NIR images for polarimetry. 

The polarization efficiencies were estimated to be 95.5\%, 96.3\%, and 98.5\% at $J$-, $H$-, and $K_{\rm s}$-bands, respectively, 
and the measurable polarization is $\sim$0.3\% over the entire field of view at each band \citep{kandori06}. 
The absolute accuracy of the polarization angle $\theta$ is estimated to be better than 3$^\circ$ (\cite{kandori06}; \cite{kusune15}). 

In figure \ref{obs_area}b, 
we show the area of our NIR polarimetric observations overlaid on the ArT\'{e}MiS 350 $\mu$m dust continuum map of NGC~6334.  
In total, we observed 6 fields toward NGC~6334 in 2023. 
For each field of view, 
we obtained 10 dithered exposures, each 15~s long, at four half-wave plate angles 
(in the sequence of 0$^\circ$, 45$^\circ$, 22.$^\circ$5, and 67.$^\circ$5 in the instrumental coordinate system) 
as one set of observations and repeated this six times. 
Eventually, 
the total on-target exposure time was 900 s per each half-wave angle. 
Self-sky images were used for median sky subtraction. 
The average seeing size ranged $\sim$1$\farcs2$ to $\sim$2$\farcs5$ at $H_{\rm s}$-band during the observations. 
Twilight flat-field images were obtained at the beginning and/or at the end of the observations. 

Standard data reduction procedures were applied with IRAF/PyRAF. 
Aperture polarimetry was performed at $J$-, $H$-, and $K_{\rm s}$-bands 
with an aperture radius of $\sim$ one FWHM of point sources by using APHOT of the DAOPHOT package. 
The Two Micron All Sky Survey (2MASS) catalog \citep{skrutskie06} was used for absolute photometric calibration. 
See Appendix for more details on the data reduction procedure, 
on the source selection, 
and on the method to derive the polarization degree $P$, 
its error $\Delta P$, 
the polarization angel $\theta$, 
and its error $\Delta \theta$.

\subsection{[C\,\emissiontype{II}] 158 $\mu$m observations} 

The [C\,\emissiontype{II}] observations of NGC~6334 were carried out during two balloon flights in 2018 and 2023 
conducted/launched from the Balloon Facility of the Tata Institute of Fundamental Research (TIFR) at Hyderabad. 
These measurements used a Japanese Fabry-P\'{e}rot spectrometer 
onboard TIFR 100-cm balloon-borne far-IR telescope (FPS100) operated 
at an altitude of $\sim$30 km in the stratosphere. 
NGC~6334 was observed with a spatially unchopped, fast spectral scan mode 
by two sets of the spatial raster scans covering an area of $\sim$20$\arcmin$ $\times$ 50$\arcmin$ (figure \ref{obs_area}b). 
The final [C\,\emissiontype{II}] map generated has a spatial resolution of 90$\arcsec$.
Details of FPS100 are presented in \citet{moo03}. 
Pointing of the telescope to NGC~6334 is achieved by offsetting it with respect to a nearby bright star, 
and an absolute pointing error is typically estimated to be $\sim$60$\arcsec$.
Calibration of the [C\,\emissiontype{II}] flux was done by comparing it to the [C\,\emissiontype{II}] map of Eta Carina \citep{mizutani04}. 
The flux calibration error is approximately 20\%.

We performed data reduction following \citet{kaneda13, suzuki21}. 
The [C\,\emissiontype{II}] intensity map was obtained by subtracting the background and continuum components from each spectral scan. 
The average spectrum in the region of $\sim$20$\arcmin$ $\times$ 7$\arcmin$, 
without [C\,\emissiontype{II}] emission in the vicinity of NGC~6334, 
was used as background. 
The spectral scan data on the corresponding spatial scans is modeled by combining the background spectrum, 
the Rayleigh-Jeans regime modified blackbody function
with the emissivity power-law index $\beta$ of 2.0 for the 158 $\mu$m continuum emission, 
and the Lorentz function  for the [C\,\emissiontype{II}] emission. 
Here, the amplitude of the background, the 158 $\mu$m continuum, and the [C\,\emissiontype{II}] emission were set to be free, 
and the center and width of the Lorentz function were fixed at the values estimated from several spectral scan data with high signal-to-noise ratios. 
To determine the positional offset of the [C\,\emissiontype{II}] map 
and to improve the positional accuracy of the [C\,\emissiontype{II}] map, 
the 158 $\mu$m continuum emission map was shifted along the equatorial coordinate axis
so that the peak of the 158 $\mu$m continuous emission map coincided with the peak of the Herschel 160 $\mu$m map. 
Finally, we regrid the two [C\,\emissiontype{II}] maps for 2018 and 2023 at a 90$\arcsec$ pixel size for pixel-by-pixel correlation analysis to obtain a composite map.

\begin{figure*}
 \begin{center}
  \includegraphics[width=8cm]{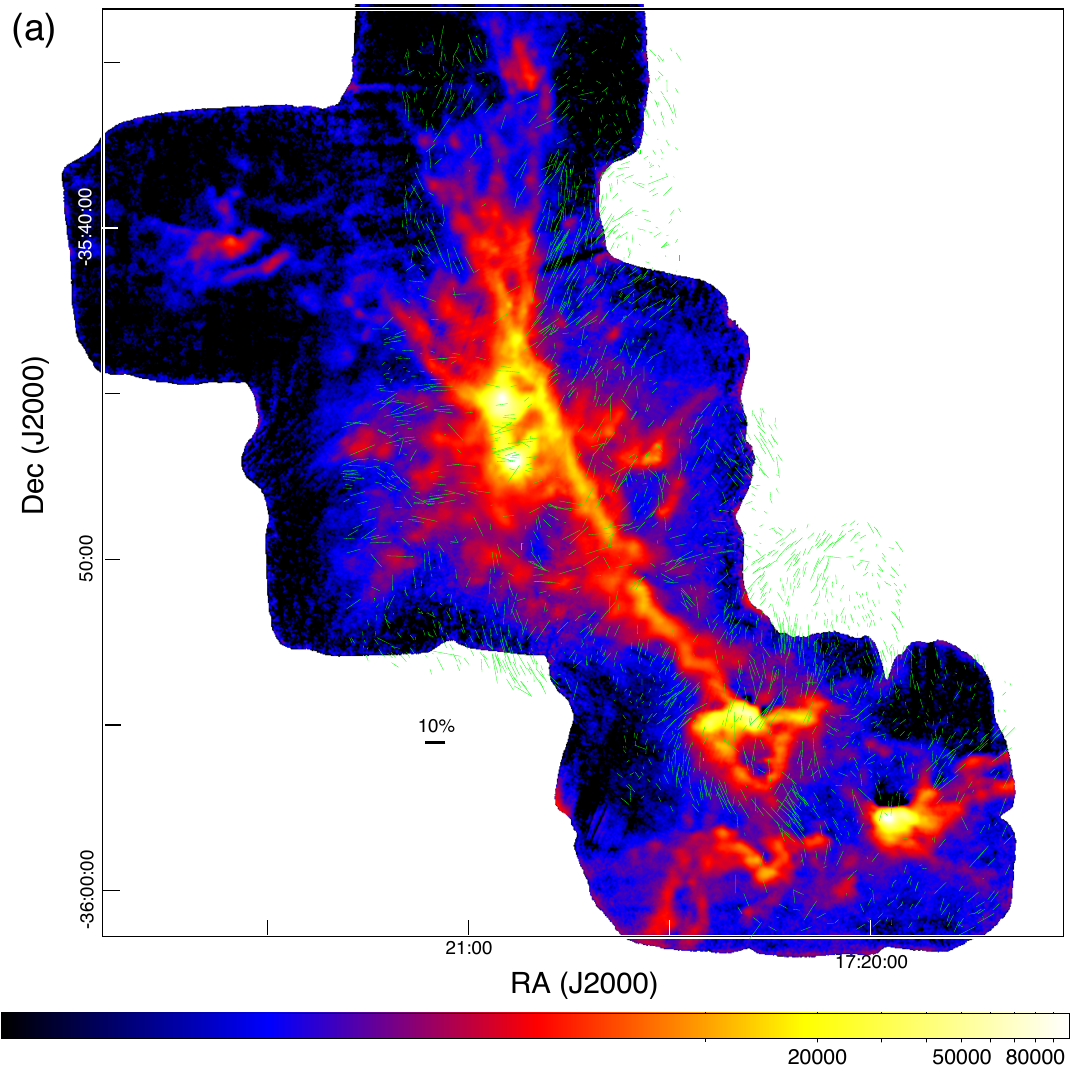}
  \includegraphics[width=8cm]{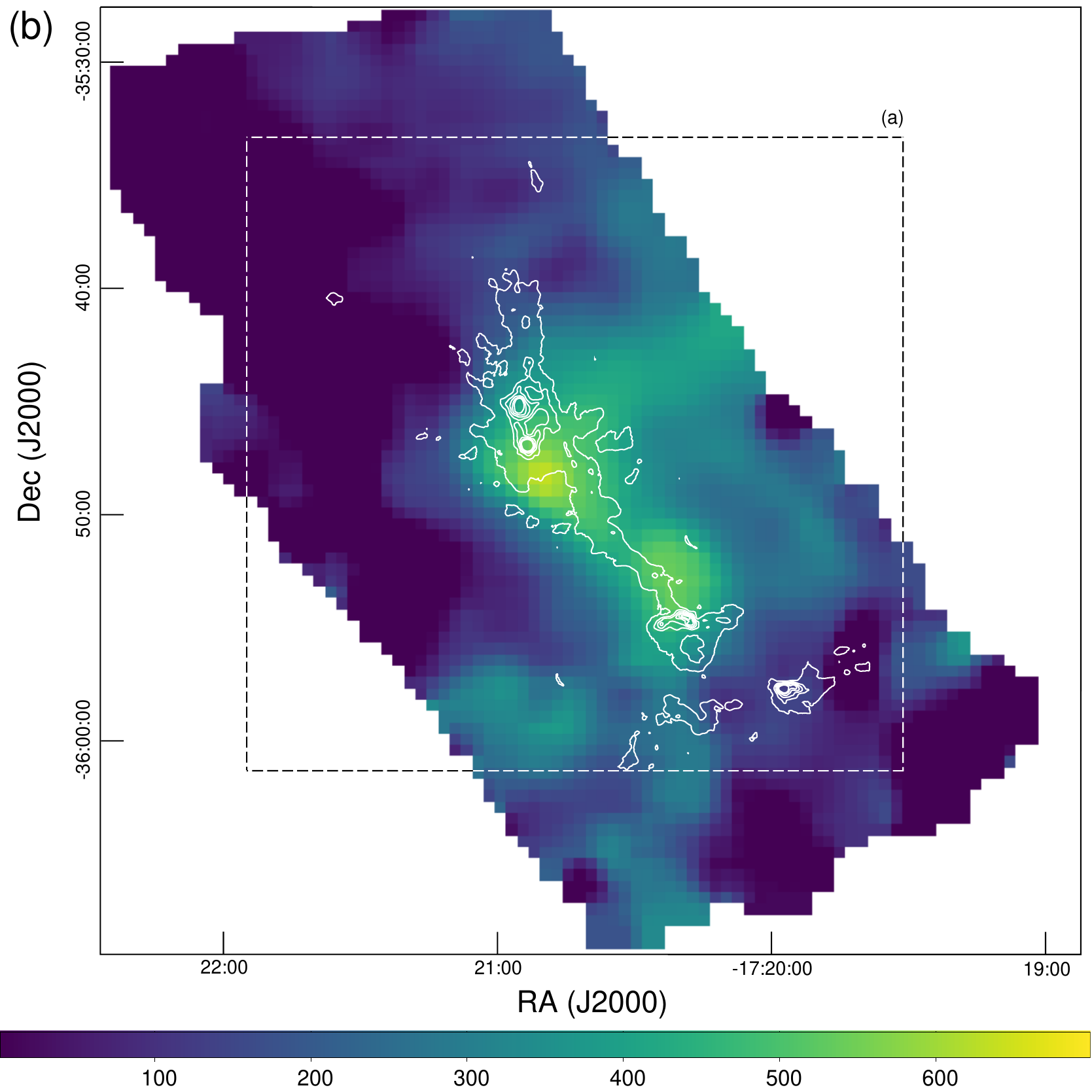}
 \end{center}
\caption{
(a) $H$-band polarization vector map of NGC~6334 
superposed on the ArT\'{e}MiS 350 $\mu$m dust continuum map at 8$''$ angular resolution from \citet{andre16} 
in unit of MJy sr$^{-1}$. 
For reference, 
the length for 10\% polarization is shown by the black vector on the left of the NGC~6334. 
(b) The [C\,\emissiontype{II}] intensity map at 90$''$ angular resolution of NGC~6334 . 
The white contours indicate the ArT\'{e}MiS 350 $\mu$m dust continuum map from \citet{andre16}. 
The dashed square indicates the area shown in figure \ref{pvm_cii}a.
The color scale gives the [C\,\emissiontype{II}] intensity in unit of 10$^{-8}$ W$^{2}$ m$^{-2}$ sr$^{-1}$.
{Alt text: A figure showing the morphology of the dust filaments and the magnetic field directions, 
and a figure showing the morphology of the dust filaments and the CII emission.}
} 
\label{pvm_cii} 
\end{figure*}

We show the $H$-band polarization vector map toward NGC~6334 overlaid on the [C\,\emissiontype{II}] map in figure \ref{pvm_cii}. 
The NIR polarization vector treated in the present paper traces the POS magnetic field of NGC~6334. 
This is because there is correlation between the polarization degree $P$ and the color $H-K_{\rm s}$ (see the Appendix), 
and the NIR polarization is mainly caused by absorption in NGC~6334 without significant absorption in the foreground of NGC~6334. 
Since the vector maps of the $J$- and $K_{\rm s}$-bands are similar to that of the $H$-band, 
we use only the $H$-band data in the following.

\subsection{Herschel and APEX/ArT\'{e}MiS archival data} 

\begin{figure*} 
\begin{center} 
\includegraphics[width=17cm]{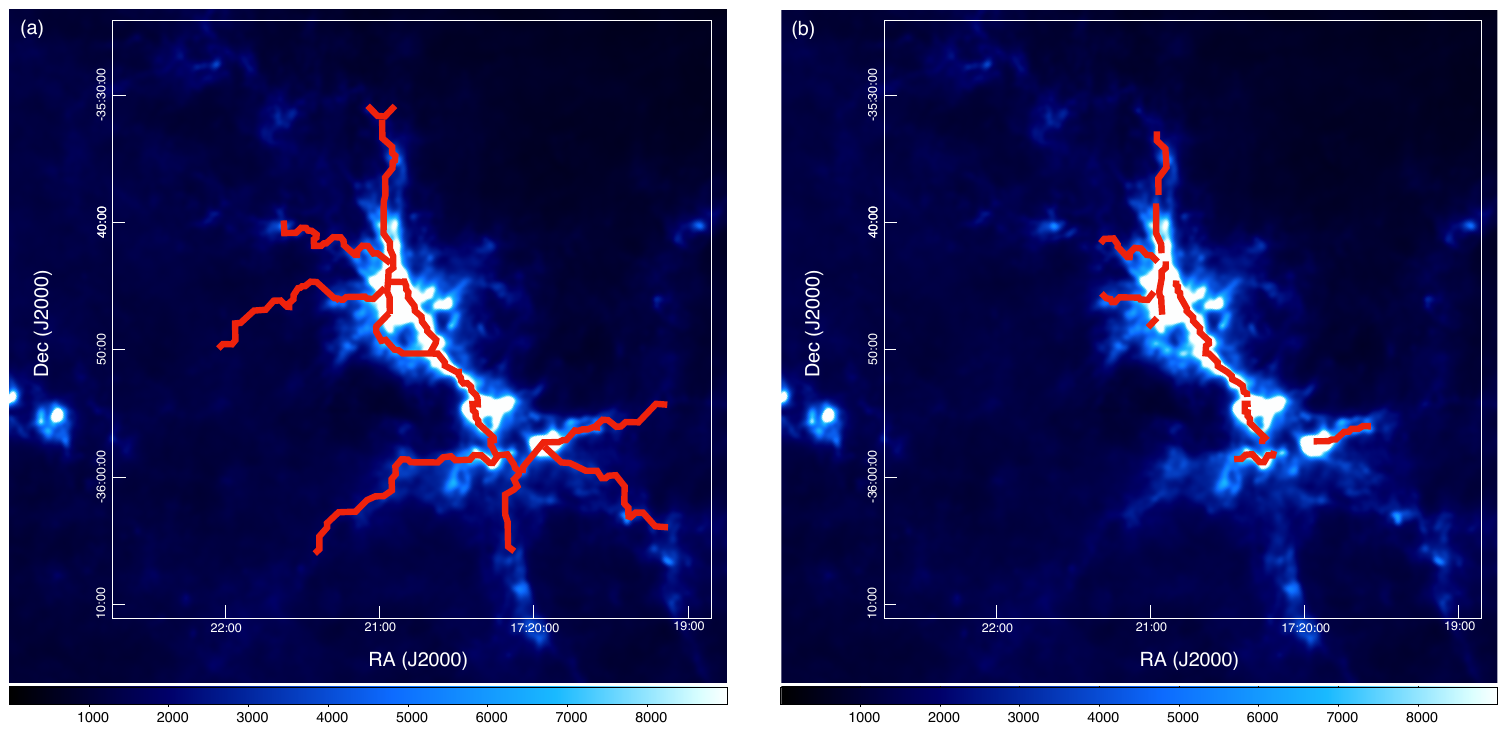} 
\end{center} 
\caption{ 
(a) Filamentary structure of NGC~6334 detected by DisPerSE. 
The background image is the Herschel 250 $\mu$m image in unit of MJy sr$^{-1}$. 
(b) Ten filaments defined in the present paper based on the results on the left panel. 
{Alt text: Two maps.} 
} 
\label{filament_disperse} 
\end{figure*}

To compare our NIR polarimetric data with the cloud structure, 
we use the Herschel Archival SPIRE 250 $\mu$m map at 18$''$ angular resolution (HOBYS; \cite{motte10}) 
and the archival APEX/ArT\'{e}MiS 350 $\mu$m dust continuum map \citep{andre16}. 
The effective angular resolution of the ArT\'{e}MiS map is 8$''$ (HPBW). 
In the present paper, 
higher-resolution ArT\'{e}MiS image is used when comparing our high-resolution POS magnetic field structure with the detailed cloud shapes, 
while Herschel image is used for the analysis of the broader field of view 
(e.g., the filament detection, the [C\,\emissiontype{II}] emission, or the PAH emission). 
In the high-resolution map shown in figure \ref{pvm_cii}a, 
fine filaments are seen connecting perpendicularly to the prominent filaments, 
and the POS magnetic field structure parallel to these fine structures is also traced.  
In order to clarify the relationship between the magnetic field, 
we define the filaments in NGC~6334. 
To trace the elongated dusty filaments seen in the Herschel 250 $\mu$m dust emission map (figure \ref{obs_area}a), 
we used the DisPerSE algorithm (\cite{sousbie11a}; \cite{sousbie11b}). 
DisPerSE works with the Morse theory and the concept of persistence 
to identify topological structures such as filaments, 
and to connect their saddle-points with maxima by integral lines. 
This algorithm has already been successfully applied to trace filamentary structures in various molecular clouds (e.g., \cite{arzou11}). 
Figure \ref{filament_disperse}a presents the filaments in NGC~6334 identified by DisPerSE. 
In the present paper, 
we divide the NGC~6334 molecular cloud into 10 filament regions based on the filaments identified by DisPerSE
to investigate the characteristics of the magnetic field, the [C\,\emissiontype{II}] emission, and the PAH emission of each filament. 
Here, 
a threshold of 1.4 sigma (i.e., -nsig=1.4 in DisPerSE) was used for filament detection to pick up the most prominent filaments.
Figure \ref{filament_disperse}b shows the 10 filaments used for the analysis. 
Note that 
we excluded the filaments outside of NGC~6334 because the NIR polarimetric and [C\,\emissiontype{II}] observations were performed around the center of NGC~6334. 
Here, although some structures were identified by DisPerSE, 
regions without accompanying dust emission were not considered as filaments.
The filamentary structures we identified using DisPerSE are roughly consistent with those identified by \citet{arzou22},
also using DisPerSE, 
although they differ somewhat in the southern region of NGC~6334.

\section{Results}

\subsection{Relationship between POS magnetic field and filaments} 

The filaments are evaluated from the point of view of the POS magnetic field direction. 
We examine the angle of the POS magnetic field associated with each filament with respect to the filament direction. 
Here, we consider a polarization vector within 1 pc of a filament to be the magnetic field associated with that filament. 
In the case of multiple filaments within 1 pc of a polarization vector, 
the vector is treated as the magnetic field associated with the nearest filament. 

\begin{figure*}
\begin{center}
\includegraphics[width=17cm]{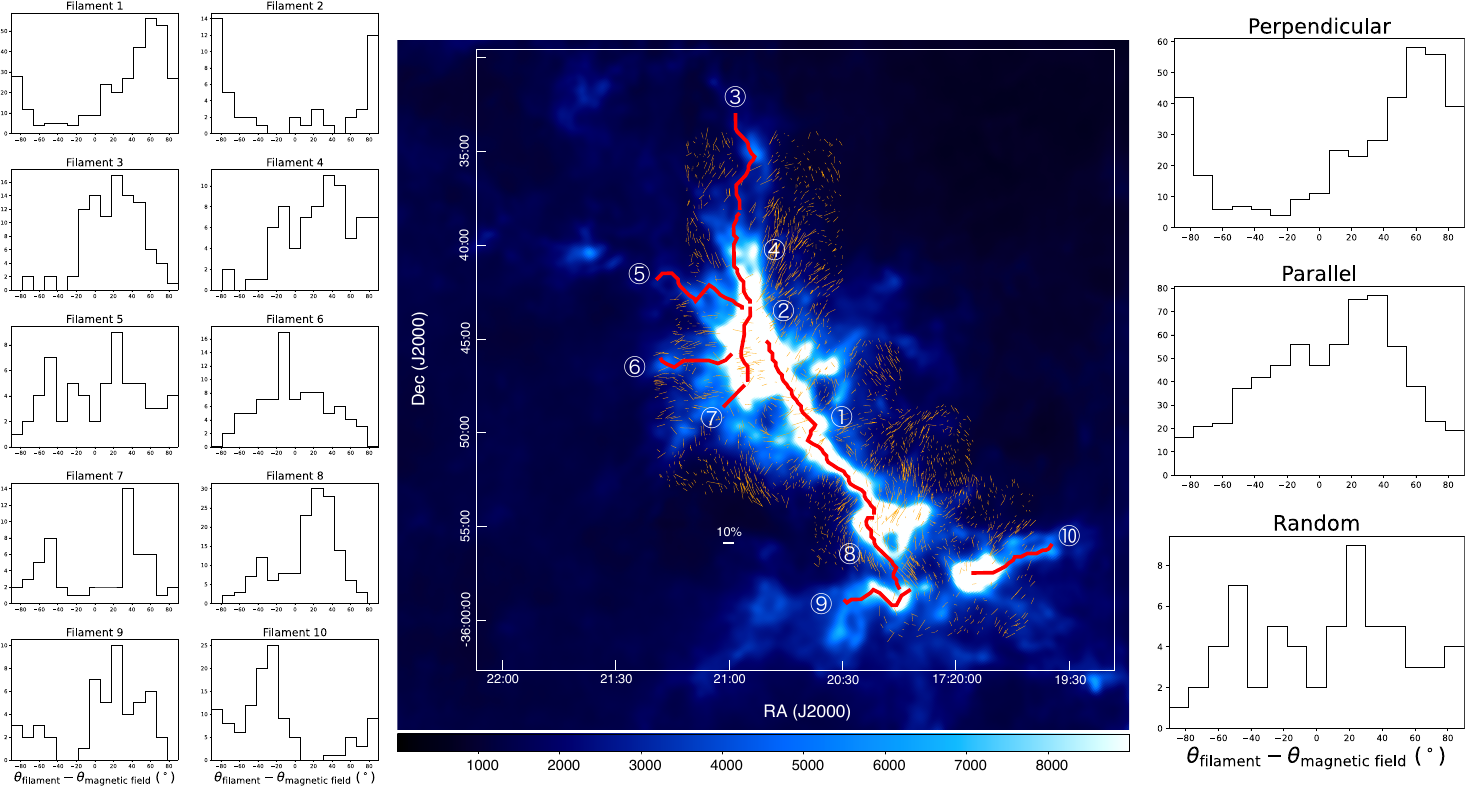} 
\end{center}
\caption{
Histograms showing the angle of the magnetic field direction with respect to the filaments' elongation for each filament. 
$\pm$90$\degree$ means that the filaments' elongation and the field direction is perpendicular, 
while 0$\degree$ means that the filaments' elongation and the field direction is parallel. 
Based on the histograms, the relationship between the filaments and field is defined as `perpendicular', `parallel', and `random'. 
The middle image is the $H$-band polarization vector with the detected 10 filaments superposed on the Herschel 250 $\mu$m image in unit of MJy sr$^{-1}$. 
The histograms on the right show stacked ones for each filament type. 
{Alt text: 
A map showing the filaments and the magnetic field directions, and histograms showing the relative angle between them. 
}
} 
\label{B-hist} 
\end{figure*}

Figure \ref{B-hist} presents a polarization vector map showing the POS magnetic field direction superposed on the dust distribution map 
and histograms showing the angle between the POS magnetic field direction and the filament direction in each filament. 
A histogram peak around 0\textdegree means that 
the POS magnetic field tends to be parallel to the filament, 
while a peak around $\pm$90$\degree$ means that 
the POS magnetic fields tends to be perpendicular to the filament. 

The present analysis shows that the NGC~6334 filaments have the following trends: 
for Filaments 1 and 2, 
which are located in the center of the NGC~6334, 
the histogram peaks around $\pm$90$\degree$, 
thus the POS magnetic field and filaments tend to be perpendicular. 
For Filaments 3, 4, 6, 7, 8, 9, and 10, 
the peaks are close to 0$\degree$, 
thus the POS field and filaments tend to be parallel. 
For the other filaments, the POS field seems to run randomly.

\subsection{Relationship between [C\,\emissiontype{II}] and PAH, and filaments} 

\begin{figure*} 
\begin{center} 
 \includegraphics[width=8cm]{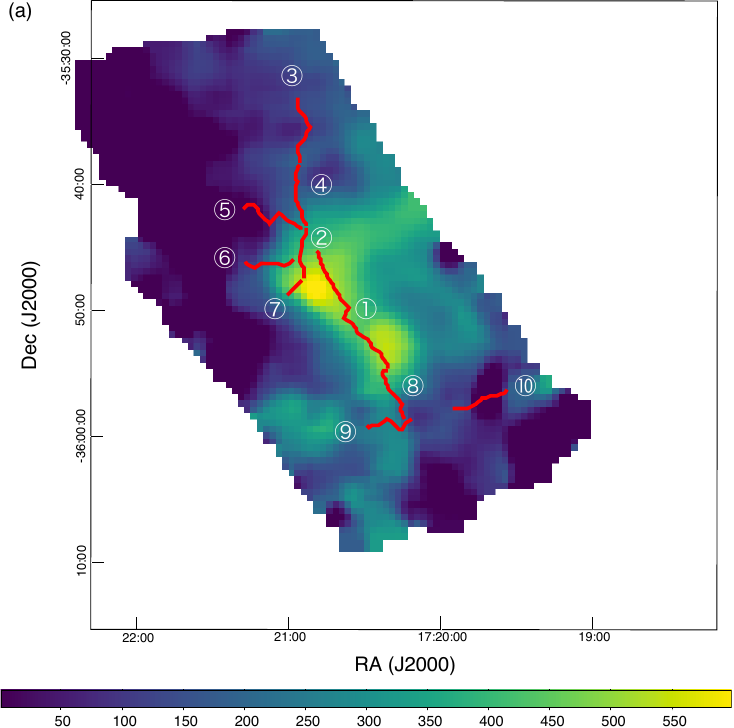}
  \includegraphics[width=8cm]{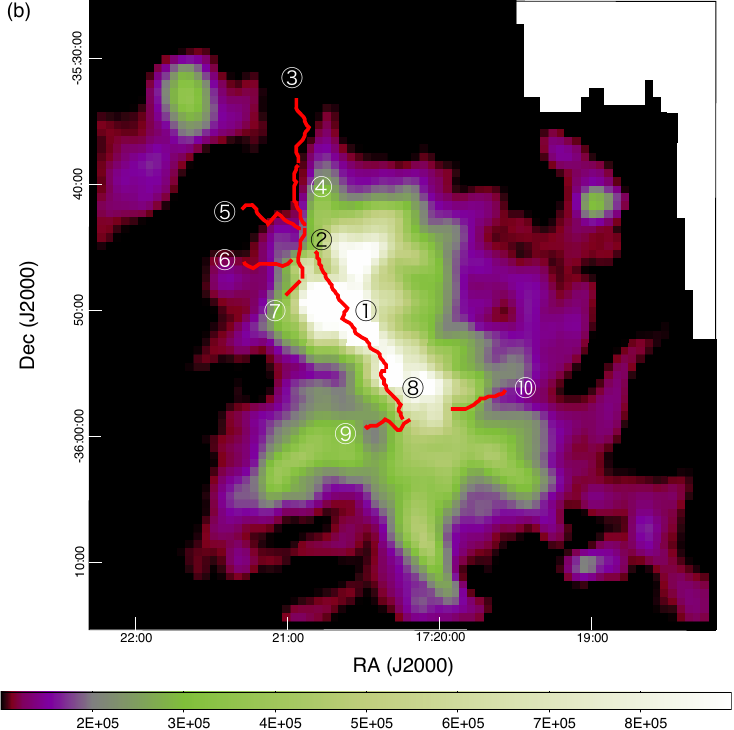} 
\end{center} 
\caption{ 
(a) The [C\,\emissiontype{II}] intensity map with the filaments detected by DisPerSE. 
The color scale gives the [C\,\emissiontype{II}] intensity in unit of 10$^{-8}$ W$^{2}$ m$^{-2}$ sr$^{-1}$.
(b) The PAH Spitzer 8 $\mu$m image overlaid with the filaments detected by DisPerSE. 
The color scale is given in unit of MJy sr$^{-1}$.
{Alt text: Two maps. }
} 
\label{cii_filament} 
\end{figure*}

Figure \ref{cii_filament}a  presents the [C\,\emissiontype{II}] intensity map with 10 filaments detected by DisPerSE. 
As shown in figure \ref{cii_filament}a, 
four of the ten filaments in NGC~6334, 
Filaments 1, 2, and 7 are the filaments with the brighter [C\,\emissiontype{II}] emission, 
indicating active star formation. 
Of these filaments, 
Filament 1 is a single filament isolated from its surroundings, 
while the other Filaments 2 and 7 form a hub filament structure. 

Figure \ref{cii_filament}b shows the Spitzer 8 $\mu$m PAH emission map with the filaments. 
Here, the Spitzer 8 $\mu$m map was smoothed to a resolution nearly matching that of the [C\,\emissiontype{II}] map for comparison. 
PAH emission, similar to [C\,\emissiontype{II}] emission (figure \ref{cii_filament}a), 
is also strong in Filaments 1, 2, and 7. 
In contrast, 
the northern filaments of NGC~6334 (Filaments 3 to 6) show weak PAH emissions, consistent with the  [C\,\emissiontype{II}] results. 
However, Filaments 8, 9, and 10, 
which do not emit strongly in  [C\,\emissiontype{II}], 
were found to emit relatively strongly in PAH.

\section{Discussion} 

\begin{figure*}
 \begin{center}
  \includegraphics[width=17cm]{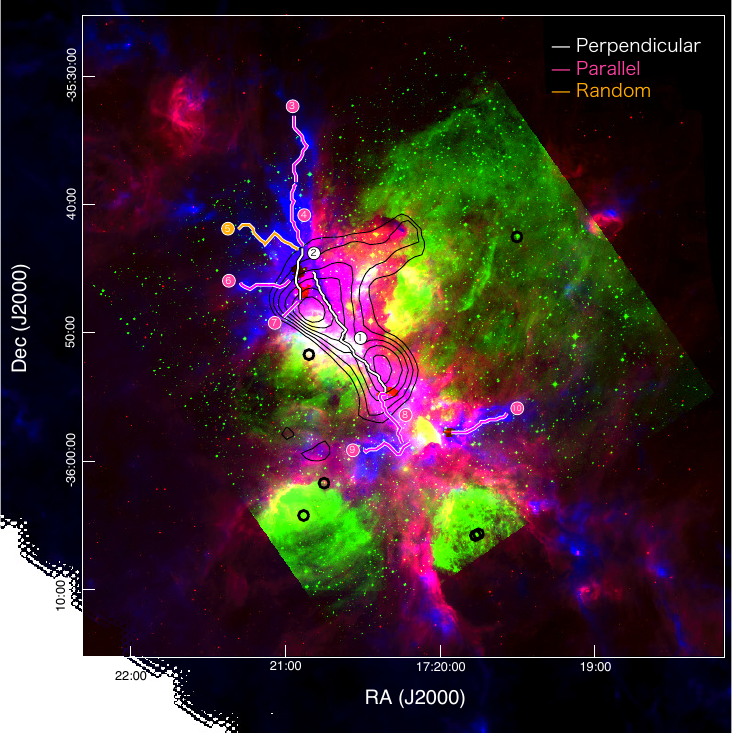} 
 \end{center} 
\caption{ 
Three-color composite image around NGC~6334 
(PAH Spitzer 8 $\mu$m image in red; H$\alpha$ image in green; Herschel 250 $\mu$m map in blue). 
The filaments detected by DisPerSE are indicated by lines: 
Filaments perpendicular to the magnetic fields, 
parallel to the magnetic fields,
random to the magnetic fields are shown in white, purple, and orange, respectively. 
The distribution of the [C\,\emissiontype{II}] intensity map is shown by the black contours.
The positions of O stars are indicated by black circles \citep{russeil20}. 
{Alt text:
A overview map showing the distribution of filaments in relation to the magnetic field direction, the locations of HII regions, and the CII emission. 
}
} 
\label{3col} 
\end{figure*}

We investigate why star formation is so active in the filaments (i.e., Filament~1) 
where the magnetic field runs vertically in NGC~6334. 
To examine the environment around NGC~6334 in more detail, 
we prepare a three-color composite image with a larger field of view in figure \ref{3col}. 
In figure \ref{3col}, 
the Spitzer 8 $\mu$m image tracing the PAH  emission is shown in red, 
the $R$-band image with the H$\alpha$ emission line in green, 
and the Herschel 250 $\mu$m map in blue.
The distribution of the [C\,\emissiontype{II}] intensity map is shown by the black contours in figure \ref{3col}.  
The positions of O stars are indicated by black circles \citep{russeil20}. 
This map shows that 
the PAH emission indicating PDR is distributed just outside the dust distribution, 
i.e., the filaments indicated by lines, 
and then the H$\alpha$ emission indicating ionized gas regions extends outside of the PAH distribution. 
In addition, the O stars are located in the center of the ionized gas regions. 
Thus, 
NGC~6334 is very likely to be associated with four H\,\emissiontype{II} regions created by O stars, 
and to be significantly affected by them.

As shown in figure \ref{3col}, 
Filament~1 appears to exist just sandwiched between the two H\,\emissiontype{II} regions. 
The magnetic field runs perpendicular to Filament~1 elongation, 
which means that 
the magnetic field is parallel to the direction of expansion of the two H\,\emissiontype{II} regions. 
Since the molecular gas and the magnetic field are frozen, 
the gas along the magnetic field can easily move. 
In Filament~1, 
such a configuration of the filament elongation, magnetic field, and the two H\,\emissiontype{II} regions 
allows the gas to accrete efficiently from the edge of the H\,\emissiontype{II} region to the filament, 
and star formation is so active that the [C\,\emissiontype{II}] emission is detected. 
\citet{arzou22} mentioned that 
the feedback from the two H\,\emissiontype{II} regions is present for Filament~1, 
which they refer to as VCF~1 or MFS-warm in their paper, 
based on the velocity gradients in the direction perpendicular to Filament~1 seen in the position-velocity diagrams for the $^{13}$CO and C$^{18}$O emission (figure 12 of \cite{arzou22}). 
On the other hand, 
Filaments~2 and 7 
form a hub filament system, 
and star formation is very likely to be enhanced by filament-to-filament interactions. 

In the case where the magnetic field and filament direction are parallel, 
it is known that the star formation in the filament tends to be inactive.  
Our observations indicate that 
the star formation is quiescent in filaments parallel to the POS magnetic field direction (Filaments 3--10, except for 5),  
and thus our results is an example of such a picture being observationally demonstrated.
Filaments~8, 9, and 10 in the southern region of NGC 6334 are filaments where PAH emission is relatively strong, while the [C\,\emissiontype{II}] emission is weak. 
This suggests that these three filaments may be more likely to be parts of the edges of H\,\emissiontype{II} regions rather than true filaments. 

It has also been reported that 
the star formation in NGC~6334 may have been caused by cloud-cloud collision \citep{fukui18}.
However, our NIR polarimetric observations show that 
the overall magnetic field structure is relatively ordered for a region 
where cloud-cloud collisions are reported to have occurred. 
In other words, 
when considering the star formation in NGC~6334, 
we may need to take into account not only collisions between giant molecular clouds, 
but also influences from surrounding H\,\emissiontype{II} regions.

\section{Conclusion}

We have reported the results of NIR polarimetric observations and the [C\,\emissiontype{II}] line observations towards NGC~6334, 
and investigated the relationship between the POS magnetic fields filament direction, and between the [C\,\emissiontype{II}] emissions and the PAH emission for each filaments. 

The $H$-band polarization map of NGC~6334 obtained from the NIR polarimetric observation shows that 
there are several different types of magnetic field directions for each filament within NGC~6334. 
In Filaments~1 and 2, 
the most prominent filaments in the center region of NGC~6334, 
the magnetic field tends to be perpendicular to the filament direction. 
On the other hand, 
for the other filaments (Filaments~3 to 10, except for 5),
the magnetic fields and the filaments tend to be in a parallel relationship. 
The [C\,\emissiontype{II}] emission distribution map reveals that 
there is a clear difference in the intensity of the [C\,\emissiontype{II}] emission for each filament:
The most prominent filament, Filament~1,
and Filaments~2 and 7, which form the hub-filament system, 
have remarkably bright [C\,\emissiontype{II}] emissions, 
meaning that 
the star formation in these filaments is extremely active.  
Similar to the [C\,\emissiontype{II}] emission map, 
the PAH emission map shows strong emission in the central part of NGC~6334 
and weak emission in the northern part. 
In contrast, Filaments~8, 9, and 10 in the southern region of NGC~6334, where [C\,\emissiontype{II}] emission is weak, 
exhibit relatively strong PAH emission. 

These observational results indicate that 
the [C\,\emissiontype{II}] emission is brighter, i.e., star formation is more active, in filaments where the POS magnetic field runs vertically. 
There are several H\,\emissiontype{II} regions surrounding NGC6334, 
and it is very likely that the H\,\emissiontype{II} regions affect star formation in the filament. 
In particular, in Filament~1, which is the [C\,\emissiontype{II}] brightest filament in NGC~6334,   
the relationship between the direction of filament elongation, 
the direction of the magnetic field, the position of the associated two H\,\emissiontype{II} regions, 
and its directions of extension, 
is thought to result in efficient accretion of material onto the filament and active star formation throughout the filament.

\begin{ack} 
We express many thanks to all the members of the Infrared Astronomy Group of TIFR and 
the members of the TIFR Balloon Facility in Hyderabad, India, 
for their support during the balloon campaign. 
We also thank the anonymous referee for improving the manuscript. 
This research is supported by a Grant-in-Aid for Scientific Research (No. 18H01252) 
from the Japan Society for the Promotion of Science (JSPS), 
the India-Japan cooperative science program (No. 120197715) 
from JSPS and the Department of Science and Technology, the Government of India. 
This research is partially 
supported by the Optical and Infrared Synergetic Telescopes for Education and Research (OISTER) program 
funded by the MEXT of Japan. 
T. K. and H. T. thank Y. Nakajima for assistance in the data reduction with the SIRPOL pipeline package. 
The IRSF project is a collaboration between Nagoya University and the South African Astronomical Observatory (SAAO) 
supported by Grants-in-Aid for Scientific Research on Priority Areas (A) (No. 10147207 and 10147214) 
and the Optical \& Near-Infrared Astronomy Inter-University Cooperation Program from MEXT of Japan 
and the National Research Foundation (NRF) of South Africa. 
DO, SKG, KS, and JPN acknowledge the support of the Department of Atomic Energy, Government of India, under project identification No. RTI 4002.
\end{ack}


\appendix 

\section{Data reduction for NIR polarimetry}

We calculated the Stokes parameters as follows: 
\begin{eqnarray} 
Q&=&I_{0}-I_{45} \\ 
U&=&I_{22.5}-I_{67.5} \\ 
I&=&\frac{1}{2}(I_{0}+I_{22.5}+I_{45}+I_{67.5})
\end{eqnarray} 
where $I_{0}$, $I_{22.5}$, $I_{45}$, and $I_{67.5}$ are intensities at four wave-plate angles. 
To obtain the Stokes parameters in the equatorial coordinate system ($Q'$ and $U'$), 
$Q$ and $U$ were rotated by 105$^\circ$ (\cite{kandori06}; \cite{kusune15}). 
We calculated the degree of polarization $P$ and the polarization angle $\theta$ as follows: 
\begin{eqnarray}
P&=&\frac{\sqrt{Q^2+U^2}}{I} \\
\theta&=&\frac{1}{2}{\rm arctan}\frac{U'}{Q'} 
\end{eqnarray}
The errors of the polarization degrees $\Delta P$  and of the polarization angles $\Delta\theta$ were calculated from the photometric errors.  
In consideration of the above measurable polarization of $\sim$0.3\%, 
for the sources with the $\Delta P<0.3$\% we adopt 0.3\% as the errors of the polarization degrees when we calculate $\Delta\theta$.

\begin{figure}
 \begin{center}
  \includegraphics[width=8cm]{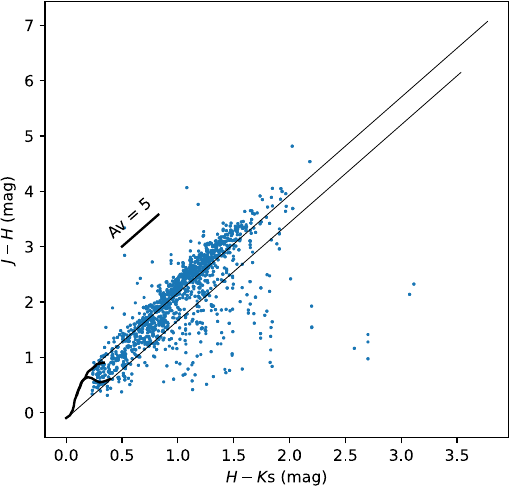} 
 \end{center}
\caption{
$J-H$ vs. $H-K_{\rm s}$ color-color diagram toward NGC~6334. 
The thick lines are the loci of dwarfs and giants. 
The two thin lines are reddening ones parallel to the reddening vector, 
of which $A_V = 5$ mag is indicated by the thick black line.
{Alt text: A diagram.}
}
\label{2cd}
\end{figure}

\begin{figure}
 \begin{center}
  \includegraphics[width=8cm]{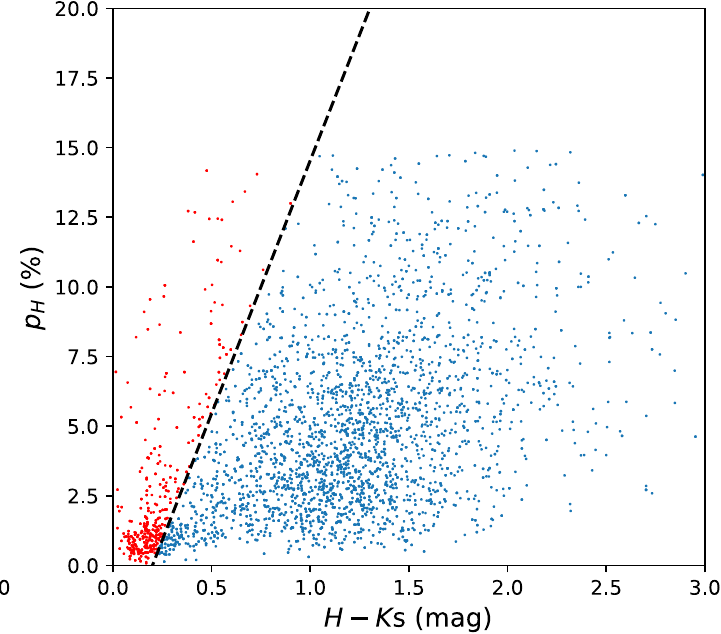} 
 \end{center}
\caption{
Polarization degree $P$ vs. $H - K_{\rm s}$ color diagram at $H$ band. 
The dashed line is the upper limit of the interstellar polarization \citep{jones89}. 
Red dots indicate the sources removed as the foreground stars.  
{Alt text: A diagram.}
}
\label{p_hk}
\end{figure}

We have measured $J$, $H$, and $K_{\rm s}$ polarizations for point sources to examine the magnetic field structure. 
We select sources for our polarization measurements using the following three procedures: 
first, 
we use only sources with photometric errors of $<$0.1 mag. 
Second, 
for the sources detected at all the bands, 
we exclude NIR sources with IR excess as YSO candidates.
They are defined as the sources plotted red-ward of the reddening line from the A0 star on the $J-H$ versus $H-K_{\rm s}$ diagram, 
which is shown in figure \ref{2cd}. 
Here we adopted the reddening line of $E(J-H)/E(H-K_{\rm s}) \sim$ 1.77 \citep{nishiyama06}. 
This value is in good agreement with our data in figure \ref{2cd}. 
Third, 
we exclude outliers whose polarization degrees significantly deviate from the interstellar polarization relation 
because the polarizations of such sources may not be caused by the dichroic process and 
therefore they are not likely to trace the magnetic field structure of a molecular cloud \citep{kusakabe08}. 
Figure \ref{p_hk} shows 
the polarization degree $P$ versus $H-K_{\rm s}$ color diagram for the point sources with $\Delta P<0.3\%$ at $H$ band.
Here we put the upper limits on the interstellar polarization \citep{jones89}, 
where we assume the intrinsic color $(H-K_{\rm s})_0 = 0.20$ of background sources 
on the basis of the model calculations by \citet{wainscoat92}. 
This upper limit is indicated as the dashed line in figure \ref{p_hk}.
In these criteria, 
the foreground sources, 
which show small polarization degrees, 
are located blue-ward of the upper limit of the interstellar polarization,
and thus are excluded from the sources used for our analysis. 
Finally, 
we use sources with $P/\Delta P>3.0$, 
which corresponds to $\Delta\theta<$ 10$^\circ$.

\end{document}